# Tailoring magnetic order via atomically stacking 3*d*/5*d* electrons

Ke Huang[1,†], Liang Wu[2,3,†], Maoyu Wang[4], Nyayabanta Swain[1], M. Motapothula[5,6], Yongzheng Luo[7], Kun Han[1], Mingfeng Chen[2], Chen Ye[1], Allen Jian Yang[1], Huan Xu[8], Dong-chen Qi[9,10], Alpha T. N'Diaye[11], Christos Panagopoulos[1], Daniel Primetzhofer[5], Lei Shen[7], Pinaki Sengupta[1], Jing Ma[2,*], Zhenxing Feng[4], Ce-Wen Nan[2,*], X. Renshaw Wang[1,8,*]

[1]Division of Physics and Applied Physics, School of Physical and Mathematical Sciences, Nanyang Technological University, 637371, Singapore

[2]School of Materials Science and Engineering, State Key Laboratory of New Ceramics and Fine Processing, Tsinghua University, Beijing 100084, China

[3]Department of Physics and Astronomy, Rutgers University, Piscataway, NJ 08854, USA

[4]School of Chemical, Biological, and Environmental Engineering, Oregon State University, Corvallis, OR 97331, USA

[5]Department of Physics and Astronomy, Uppsala University, Box 516, SE-75120 Uppsala, Sweden

[6]Department of Physics, SRM University AP, Amaravati, Andhra Pradesh 522-502, India[7]Department of Mechanical Engineering, National University of Singapore, 117575, Singapore

[8]School of Electrical and Electronic Engineering, Nanyang Technological

University, 639798, Singapore

[9]ARC Centre of Excellence in Future Low-Energy Electronics Technologies, School of Chemistry, Physics, and Mechanical Engineering, Queensland University of Technology, Brisbane, Queensland 4001, Australia.

[10]Department of Chemistry and Physics, La Trobe Institute for Molecular Science, La Trobe University, Melbourne, Victoria 3086, Australia.

[11]Advanced Light Source, Lawrence Berkeley National Laboratory, Berkeley, California 94720, USA

[†]These two authors contribute equally.

*Corresponding Email: renshaw@ntu.edu.sg, cwnan@mail.tsinghua.edu.cn & ma-jing@tsinghua.edu.cn

*Abstract*

The ability to tune magnetic orders, such as magnetic anisotropy and topological spin texture, is desired in order to achieve high-performance spintronic devices. A recent strategy has been to employ interfacial engineering techniques, such as the introduction of spin-correlated interfacial coupling, to tailor magnetic orders and achieve novel magnetic properties. We chose a unique polar-nonpolar $LaMnO_3/SrIrO_3$ superlattice because Mn (3d)/Ir (5d) oxides exhibit rich magnetic behaviors and strong spin-orbit coupling through the entanglement of their 3d and 5d electrons. Through magnetization and magnetotransport measurements, we found that the magnetic order is interface-dominated as the superlattice period is decreased. We were able to then effectively modify the magnetization, tilt of the ferromagnetic easy axis, and symmetry transition of the anisotropic magnetoresistance of the $LaMnO_3/SrIrO_3$ superlattice by introducing additional Mn (3d) and Ir (5d) interfaces. Further investigations using in-depth first-principles calculations and numerical simulations revealed that these magnetic behaviors could be understood by the 3d/5d electron correlation and Rashba spin-orbit coupling. The results reported here demonstrate a new route to synchronously engineer magnetic properties through the atomic stacking of different electrons, contributing to future applications.

**Introduction**

In the pursuit of unexpected functionalities and high-performance spintronic devices, intensive efforts have been devoted to the controlling of the magnetism of magnetic materials. Among various possible controlling approaches, interfacial engineering promises accessibility to not only the magnetic modulation but also a strong interfacial coupling, arousing emergent phenomena and applications[1,2]. Recently, spurred by the unprecedented advances in the material fabrication[3], the interface-based method now becomes feasible.

The key to the design of a controllable magnetism through the interface is to introduce a spin-correlated interfacial coupling. One of the most tantalizing choices is the coupling between 3*d* and 5*d* electrons. The heterostructure of Mn (3*d*)/Ir (5*d*) oxides is particularly attractive because the entanglement of 3*d* and 5*d* electrons exhibit rich magnetic behaviors[4–6] and strong SOC[7–9]. For instance, emergent ferromagnetism and topological Hall effect have been reported in the $SrMn^{4+}O_3/SrIr^{4+}O_3$[10–12] (SMO/SIO) and $La_{0.7}Sr_{0.3}MnO_3$ (LSMO)/SIO[13,14] heterostructure, respectively. Moreover, through atomically control of the coupling between the iridate and manganite, a tunable in-plane ferromagnetic anisotropy characterized by both magnetization and anisotropic magnetoresistance (AMR) was reported at the $La_{1-x}Sr_xMnO_3$/SIO[15,16] heterostructures. Up to now, while interesting results have been achieved at the $Mn^{4+}/Ir^{4+}$ or $Mn^{3+/4+}/Ir^{4+}$ interfaces, little study explores the perovskite interfaces of pure $Mn^{3+}$ and $Ir^{4+}$, which has a unique polar/nonpolar structure. With the $Mn^{3+}/Ir^{4+}$ interface provides an extra polar discontinuity at the interface[17], a stronger interfacial coupling is expected, possibly leading to a dramatic modulation of magnetism.

In this work, through atomically changing the stacking of *3d/5d* electrons, we achieved the magnetic-order modulation in $LaMn^{3+}O_3$ (LMO)/SIO superlattices, which manifests as a change of magnetization, rotation of ferromagnetic easy axis, and transition of AMR symmetry. Both magnetization and magnetotransport measurements show that the magnetic order becomes more interface-dominated while we decrease the superlattice period. Further comparison with the single-layer LMO and SIO and in-depth theoretical prediction demonstrate that the underlying mechanism for these behaviors is related to the Mn-Ir coupling and the interfacial Rashba effect.

**Results and discussion**

Four superlattice samples [(SIO)$_m$-(LMO)$_m$]$_n$, namely [(SIO)$_8$-(LMO)$_8$]$_1$, [(SIO)$_4$-(LMO)$_4$]$_2$, [(SIO)$_2$-(LMO)$_2$]$_4$ and [(SIO)$_1$-(LMO)$_1$]$_8$ (denoted as SL881, SL442, SL224 and SL118 for short) were epitaxially grown on (001)-orientated SrTiO$_3$ (STO) substrates by pulsed laser deposition at 700 °C and 100 mTorr oxygen partial pressure. Figure 1a shows the schematic structures of the superlattices. The 3$d$/5$d$ interfacial effect is controlled via atomically change the repeating period (2$m$). From SL881 to SL118, the period was shortened while the total thickness of each constituent ($m$ x $n$) was fixed. Thus, as the period is shortened, more interfaces are created, and a larger interfacial effect is expected.

The high structural quality of the superlattice samples was confirmed by the *in-situ* reflection high-energy electron diffraction (RHEED) (Fig. 1b) and X-ray diffraction (XRD) (Fig. 1c). Every oscillation in Fig. 1b represents the epitaxial growth of one unit cell (uc) LMO or SIO. The periodic oscillation of the four superlattice samples indicates a layer-by-layer growth mode. Figure 1c shows the XRD (002)-diffraction of the four superlattices. Albeit our samples are very thin, the first order superlattice satellite peaks can still be observed, demonstrating a good crystallinity. Figure 1d shows the temperature-dependent sheet resistance ($R_S$) of the four superlattices. All superlattices exhibit a semiconducting behavior with their $R_S$ falling between the conducting 8 uc SIO and insulating LMO. Compared with the extremely insulating LMO and 2 uc SIO, the measurable $R_S$ of the low-repeating-period superlattice indicates that there is an interfacial effect which could greatly modulate the transport. Moreover, SL881 has only one interface, therefore, the relatively weak interfacial effect does not impose a significant impact on its transport compared with the other superlattices.

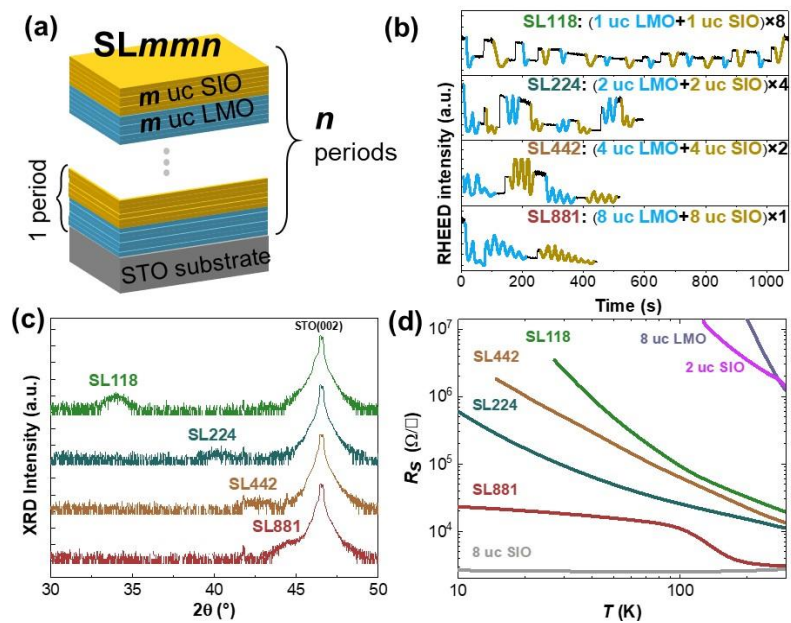

Figure 1. Basic characterization of the 3*d*/5*d* superlattice. (a) Schematic structure, (b) *in-situ* reflection high-energy electron diffraction (RHEED)

oscillations confirming the structural quality of the superlattice samples, (c) XRD patterns of the superlattice samples, and (d) temperature-dependent sheet resistance, $R_S$, of the superlattices, single-layer SIO, and LMO.

To further investigate the nature of the interfacial effect, we studied the magnetism of the four superlattices in detail. Figure 2a shows the (100)-direction temperature-dependent magnetization measured with superconducting quantum interference device (SQUID). The measurement is conducted during warming the samples with 0.1 T external magnetic field after cooling them in a 1 T magnetic field. The result shows a decreasing magnetization ($M_S$) of the superlattice when the repeating period is shorter. To be noted, when we replace the SIO with STO, the magnetization of $[(STO)_1$-$(LMO)_1]_8$ is becomes much smaller than the one of SL118 which indicates the interfacial effect is likely to be the interfacial coupling between Ir and Mn. Then, we investigate the magnetic order through the measurement of the in-plane field-dependent magnetization ($MH$) at 10 K (Fig. 2b). The hysteresis loop reveals the ferromagnetic nature of all four superlattices. More importantly, the disappearing hysteresis of the reference sample $[(STO)_1$-$(LMO)_1]_8$ indicates the Ir plays a crucial role for the ferromagnetism. Quantitative comparison of saturation $M_S$ and coercivity ($H_C$), summarized in Fig. 2c, demonstrates that both $H_C$ and $M_S$ can be tailored more than 100% from SL881 to SL118. This also shows the feasibility of tuning the magnetism through Ir-Mn interfacial engineering. To understand this behavior, we should first find out the source of the superlattice ferromagnetism. Previous work has reported Ir-moment is much weaker than Mn-moment[15]. Therefore, the magnetization measurement is dominated by Mn. LMO is an antiferromagnetic material[18] and will show ferromagnetism when it is in thin-film form through an intrinsic electron accumulation called "polar catastrophe"[4,5]. When the polar LMO epitaxial grown on a nonpolar substrate like STO, there will be a potential build-up that can drive the electrons of LMO to accumulate at the bottom. When the thickness of the LMO becomes larger than 3 uc, this accumulation can initiate a phase transition of LMO from antiferromagnetism to ferromagnetism[5]. This theory well explains the absence of ferromagnetism in the reference sample [(STO)$_1$-(LMO)$_1$]$_8$ because the thickness of individual LMO is below 3 uc. However, with the same structure by simply replacing the Ti with Ir, SL118 shows ferromagnetism. Such a transition not only confirms the existence of the interfacial coupling between SIO and LMO, but also proves the interfacial coupling is the source for the ferromagnetism in SL118.

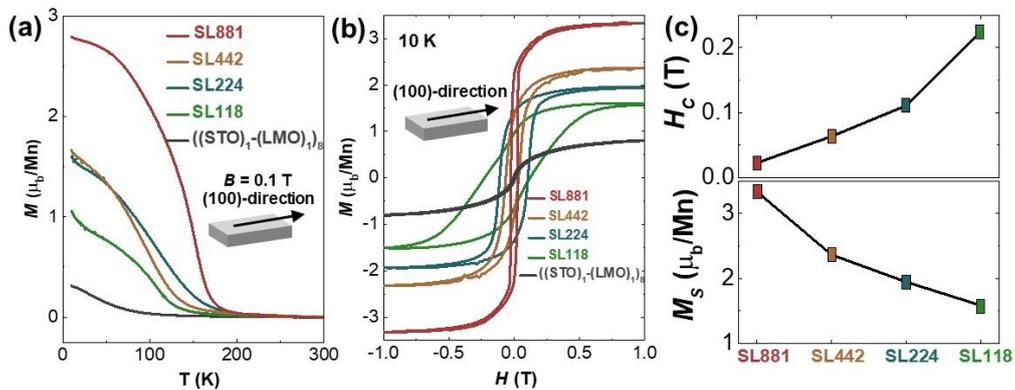

Figure 2. Magnetization of the superlattices. (a) (100)-direction temperature-dependent magnetization measured 0.1 T external magnetic field. (b) (100)-direction *MH* measured at 10 K. A reference sample [(STO)$_1$-(LMO)$_1$]$_8$ is also presented to (c) The changing of $H_C$ and $M_S$ summarized from the *MH* measurement.

Since the electronic structure of Mn decides the underlying mechanism for the magnetism, we studied the electronic state of Mn through the Mn $L_{2,3}$ edges X-ray absorption spectroscopy (XAS) measured at 85 K (Fig. 3a). The XAS of 2 and 8 uc LMO are also presented here for reference. Consistent with the previous report, the $Mn^{2+}$ spectral peak is observed in our single-layer LMO due to the electron accumulation[5]. Same as the 8 uc LMO, the XAS of SL881 also shows an $Mn^{3+}$ spectral peak together with a hump at the $Mn^{2+}$ position. This agrees with our previous result that SL881 has relatively weak interfacial coupling and its electronic structure of Mn is mainly dominated by the LMO bulk part. However, with more Ir-Mn created, the $Mn^{2+}$ peak disappears. A reasonable explanation for this would be the intrinsic electron accumulation in LMO is suppressed or counteracted. Because when the repeating period drops, the polar LMO layers will be continuously separated by nonpolar SIO and forming a nonpolar/polar/nonpolar structure. Therefore, there would be opposite polar discontinuities at the top and bottom of the LMO, and the electron accumulation along the two directions will reduce each other. Since the electron accumulation is gradually modulated from SL881 to SL118, the underlying mechanism for the ferromagnetism also undergoes a transition from the intrinsic charge transfer inside LMO to the emergent electronic correlation at the 3*d*/5*d* interface.

Apart from the modulation of the magnetization, we also achieved the tilt of the ferromagnetic easy axis. Figure 3b and 3c show the Mn *L* edge X-ray magnetic circular dichroism (XMCD) spectra[19] measured at 85 K under 1 T magnetic field with 45° and 90° (out-of-plane) incident X-rays, respectively. The magnetic field applied is in the same direction as the X-ray. Note that the out-of-plane signal is magnified 5 or 10 times for a better comparison. Figure 3b shows that all four superlattices have almost the same moment along 45° direction. Since our SQUID result (Fig. 2b) has shown that the in-plane magnetization is gradually decreasing when there are more interfaces. As a result, the spins of the four superlattices must be oriented differently, otherwise, it is impossible to have a similar net moment along the 45° direction. The 90° XMCD (Fig. 3c) further validates this conclusion. There is no out-of-plane XMCD of SL881, indicating the spins are fully in the in-plane direction. When we shorten the superlattice period from SL881 to SL118, we clearly observe an increasing out-of-plane XMCD signal indicating the ferromagnetic easy-axis tilts away from the in-plane.

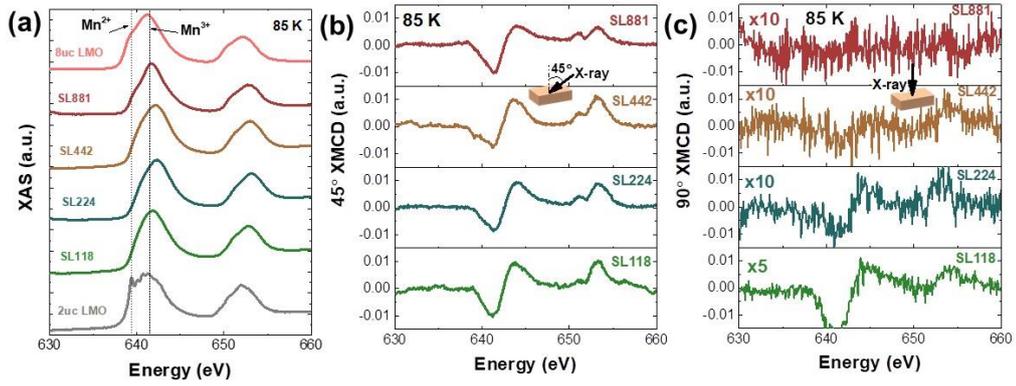

Figure 3. Electronic structure of the superlattices measured through XAS and XMCD at 85 K. (a) XAS of the superlattices. XAS of 2 uc and 8 uc LMO are presented as references. XMCD of the superlattice with different X-ray incident angle (b) 45° incident (c) 90° incident (OOP). Note that the 90° XMCD of SL118 is magnified 5 times while the others are magnified 10 times for comparison with the 45° XMCD.

Besides the ferromagnetism, we also investigate the interfacial coupling through magnetotransport study. Figure 4a shows the (100)-direction magnetoresistance (MR) measured at 30 K. SL881 and 8 uc SIO share a similar MR. This is expected because the interfacial effect of SL881 is relatively weak, and the conducting is dominated by the 8uc SIO. As the superlattice period is shortened, there is a marked increase of MR magnitude. Since the 2uc SIO or LMO are expected to be completely insulating at 30 K (Fig. 1d), their resistance will be independent of the external field. Consequently, this marked MR increase is not a pure behavior of SIO. A high-temperature MR comparison is shown in Supplementary Figure 1, which confirms this conclusion. Therefore, there should be an interfacial coupling between the SIO and LMO, which influences the magnetotransport behavior of the superlattice. The inset in Fig. 4a shows the MR hysteresis loop, which is caused by the ferromagnetic spin scattering. Moreover, the MR hysteresis loop shows the same $H_C$ as the one from the *MH* loop (Supplementary Figure 2), which proves that the spin scattering comes from ferromagnetic LMO. As a result, when the superlattice period is shortened, the transport property of the superlattice shifts from bulk SIO dominating to the LMO/SIO interface dominating, where both LMO and SIO contribute.

More information on the interfacial transport is revealed through investigating the anisotropy of the magnetotransport, we found the AMR symmetry could be tailored through changing the repeating period. The AMR was measured through rotating the in-plane 9 T magnetic field and calculated through

$$AMR(\theta) = \frac{\rho[B(\theta)] - \rho[B(\theta=0)]}{\rho[B(\theta=0)]}, \quad (1)$$

where *θ* represents the angle between the magnetic field and the fixed current direction. Figure 4b shows the experimental AMR and its fitting by the phenomenological equation[20,21]

$$R(\theta) = A_0 + A_2 \sin[2(\theta - \theta_1)] + A_4 \sin[4(\theta - \theta_2)], \quad (2)$$

where $A_2$ and $A_4$ denote the coefficient of the two-fold and four-fold symmetric term, respectively. Note that the AMR of SL881 is magnified 10 times and still too noisy to be fitted. Figure 3c shows the fitting results, which are given in the form of $A_4/A_2$. When the repeating period is shortened, the AMR gradually shifts from four-fold symmetry dominating ($A_4/A_2 >1$) to two-fold symmetry dominating ($A_4/A_2 <1$). The SL224, which is the intermediate state, has an AMR lies in between these two limits. The temperature-dependent AMR transition can be found in Supplementary Figure 3.

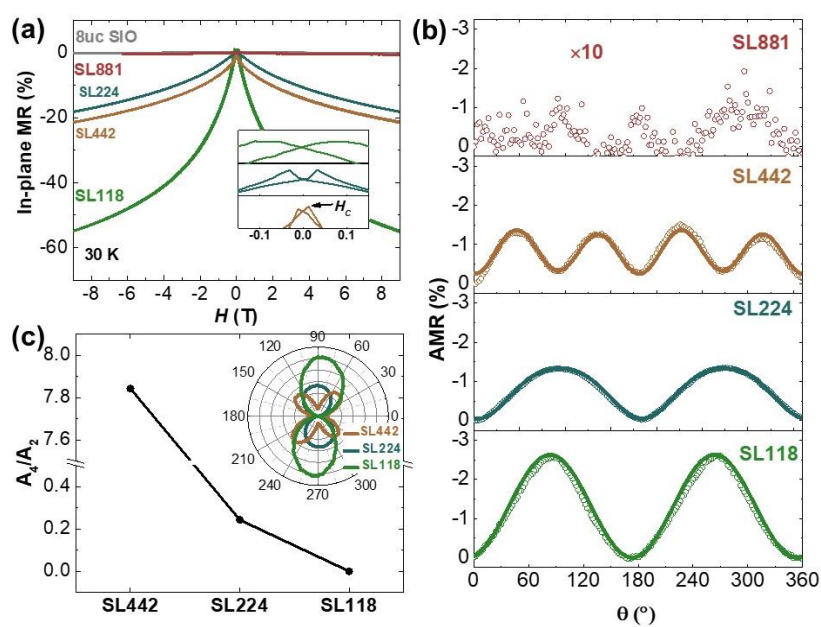

Figure 4. Evolution of magnetotransport. (a) In-plane magnetoresistance (MR) of four superlattices and 8 uc SIO. The inset shows the MR hysteresis loop, indicating the LMO contributes to the conducting. (b) In-plane anisotropic magnetoresistance (AMR) of four superlattices with the in-plane magnetic field rotating from 0° to 360°. The empty circles are the experimental results, and the solid lines represent the fitting curves. (c) The fitting results of AMR presented in an $A_4/A_2$ form.

Previous work reported that the AMR symmetry of La$_{2/3}$Sr$_{1/3}$MnO$_3$ (LSMO)/SIO superlattice is dominated by the ferromagnetic anisotropy of the LSMO[15]. However, different from the LSMO, LMO is insulating, and the ferromagnetic order of LMO/SIO superlattice is highly affected by the interfacial electronic correlation. Therefore, the explaination of the AMR transition requires the interfacial coupling. Since 9 T is much larger than the $H_C$ of the superlattices, all the spins are fully polarized along the external magnetic field direction. The symmetry of the AMR represents the intrinsic anisotropy induced by spin scattering[22], and its transition corresponds to the change of the ferromagnetic order symmetry. To explain the relation between the magnetic scattering and the AMR symmetry transition, the Rashba SOC[23,24], a type of SOC in an inversion asymmetric structure, should be considered.

We used the first-principles calculation to model the LMO/SIO superlattices with a periodic boundary condition. Figure 5a shows a representative schematic of an SL22. More details of the modeling can be found in the Supplementary Material. We obtain the Rashba SOC energy ($E_{SOC\_Rashba}$) by excluding the intrinsic SOC energy ($E_{SOC\_intrinsic}$) from the total SOC energy ($E_{SOC}$), *i.e.*, $E_{SOC\_Rashba} = E_{SOC} - E_{SOC\_intrinsic}$. By using the density functional theory (DFT) calculation embedded in the Vienna *ab initio* Simulation Package (VASP), we calculated the $E_{SOC}$ of the superlattices, bulk LMO, and SIO. The intrinsic SOC originates from the atomic ortital and is independent of the symmetry. Then, using the isolated LMO and SIO as the references, we obtain the $E_{\text{SOC\_intrinsic}}$ of the superlattices as the average of the $E_{\text{SOC}}$ of the bulk LMO and SIO, given in the form of $E_{\text{SOC\_instrinsic}} = \frac{1}{2}(E_{\text{SOC\_LMO}} + E_{\text{SOC\_SIO}})$. Figure 5b shows that the Rashba SOC is enlarged 5 times when the repeating period is shortened from SL44 to SL11. The enhancement of Rashba effect is interesting and may come from various reasons, such as the asymmetric interfacial structre[15,25,26] and built-in electric field[27]. On the basis of our first-principles calculation results, we hypothesis the enhancement of the Rashba effect comes from the asymmetric interfacial structre. Our result shows that the bond length of interfacial O-Ir-O (Fig. 5a) is monotonically decreasing from SL44 to SL11 (Fig. 5b). This decreased bond length enlarges the overlaps between different orbitals and can result in a larger gradient of internal electric potential ($\nabla V$), and therefore a larger Rahsba SOC energy.

To investigate how the Rashba SOC affects the magnetotransport properties, we numerically simulate the AMR behavior based on the first-principles results. We modeled the magnetotransport properties of the electrons at the LMO/SIO interface with a generalized Kondo lattice described by the Hamiltonian[28,29]

$$\hat{H} = -t\sum_{\langle i,j\rangle,\sigma} c_{i\sigma}^{\dagger} c_{j\sigma} + i\lambda_R \sum_{\langle i,j\rangle,\sigma\sigma'} c_{i\sigma}^{\dagger} \left(\vec{\sigma} \times \hat{r}_{ij}\right)_{z,\sigma\sigma'} c_{j\sigma'} - \mu \sum_i n_i - J_K \sum_i \boldsymbol{S}_i \cdot \boldsymbol{s}_i - B \sum_i \hat{\boldsymbol{n}} \cdot \boldsymbol{S}_i \ , \quad (3)$$

where $t$ is the hopping amplitude associated with electron motion, $\lambda_R$ is the Rashba SOC, and $J_K$ is the Kondo interaction between the Mn moments and the itinerant electron, $B$ is the strength of the external magnetic field, and $\mu$ controls the electron density. A detailed elaboration of the Hamiltonian can be found in the Supplementary Material. When the repeating period is shortened, the $J_K$ increases accordingly. The longitudinal conductivity, $\sigma_{xx}$, is calculated with the Kubo formula[30],

$$\sigma_{xx} = \frac{2\pi e^2}{Nh} \sum_{m,n \neq m} \left(\frac{f_m - f_n}{\varepsilon_m - \varepsilon_n}\right) \frac{\eta |\langle m|J_x|n\rangle|^2}{(\varepsilon_m - \varepsilon_n)^2 + \eta^2}. \quad (4)$$

Our result shows $\sigma_{xy} \approx 0$ (see Supplementary Note 5). Therefore the longitudinal resistivity is computed as $\rho_{xx} = \frac{\sigma_{xx}}{\sigma_{xx}^2 + \sigma_{xy}^2} \approx 1/\sigma_{xx}$. Figure 5c shows the theoretically-predicted AMR behavior. When we enhance the SOC term $\lambda_R$, the dominating symmetry of the AMR changes from four-fold to two-fold, which is qualitatively in agreement with the experimental observation. This simulation result proves that the coupling between the spin-orbit coupled state of $Ir^{4+}$ could impact the symmetry of the ferromagnetic order of $Mn^{3+}$ when their interfacial coupling is prominent.

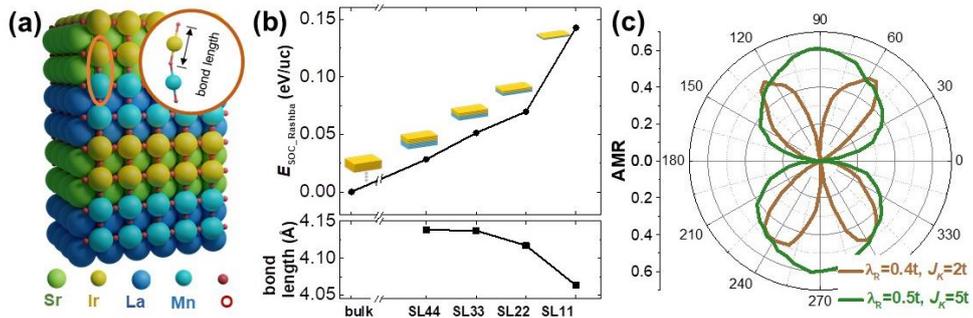

Figure 5. Theoretical investigation of LMO/SIO superlattices. (a) Schematic of the LMO/SIO superlattice structure used in the first-principle calculation. (b) Bond lengths and Rashba SOC energies obtained by first-principles calculation.

(c) Numerical simulation of anisotropic magnetoresistance (AMR) with different Rashba SOC strength.

**Conclusions**

We have observed emergent ferromagnetism in all LMO/SIO superlattices with any period even when the thickness of the LMO is below the critical thickness for polar catastrophe. Through systematically decreasing the repeating period, we achieved the modulation of the magnetization, tilt of the easy axis, and transition of AMR symmetry. The evolution of these phenomena is attributed to the *3d/5d* interface electronic correlation and the Rashba SOC. Remarkably, the interface sensitivity enables precise control of the ferromagnetism and is highly exploitable in future spintronic devices where different magnetic functionalities are simultaneously controlled.

**Supplementary Material**

See "Supplementary Material" for more details on the DFT calculation and numerical simulation.

**Acknowledgment**

X.R.W. acknowledges supports from the Nanyang Assistant Professorship grant from Nanyang Technological University and Academic Research Fund Tier 1 (RG108/17 and RG177/18) and Tier 3 (MOE2018-T3-1-002) from the Singapore Ministry of Education. S.L. acknowledges the support from Singapore MOE Tier 1 (Grant R-265-000-615-114). N.S. and P.S. acknowledge the support of MOE2014-T2-2-112 from the Ministry of Education, Singapore, and the computational resources of NSCC ASPIRE1 cluster in Singapore. D.Q. acknowledges the support of the Australian Research Council (Grant No. FT160100207). C.P. acknowledges financial support from Academic Research Fund Tier 3 (Reference No. MOE5093) and the National Research Foundation (Reference No. NRF-NRFI2015-04). C.W.N. acknowledges the supports from Basic Science Center Program of NSFC (Grant No. 51788104). Part of this research was undertaken on the soft X-ray spectroscopy beamline at the Australian Synchrotron, part of ANSTO , and the rest soft X-ray spectroscopy measurements were performed at beamline 6.3.1 of Advanced Light Source, which is an Office of Science User Facility operated for the U.S. DOE Office of Science by Lawrence Berkeley National Laboratory and supported by the DOE under Contract No. DEAC02-05CH11231.